\documentclass[12pt]{article}
\usepackage{epsfig}
\usepackage{lineno}
\usepackage{ifthen}
\newboolean{articletitles}
\setboolean{articletitles}{true} 

\usepackage{mciteplus}

\def\Journal#1#2#3#4{{\em #1} {\bf #2}, #3 (#4)}

\def\beq{\begin{equation}}
\def\eeq#1{\label{#1}\end{equation}}
\def\eeqn{\end{equation}}

\def\beqa{\begin{eqnarray}}
\def\eeqa#1{\label{#1}\end{eqnarray}}
\def\eeqan{\end{eqnarray}}

\let\bar=\overbar

\def\Dslash{\not{\hbox{\kern-4pt $D$}}}
\def\dslash{\not{\hbox{\kern-2pt $\del$}}}

\def\msb{{\bar{\ssstyle M \kern -1pt S}}}

\def\Title#1{\begin{center} {\Large {\bf #1} } \end{center}}

\begin{document}
\linenumbers

\Title{Measurement of time-dependent $\mathcal{CP}$ violation in charmless $B$ decays at LHCb}

\bigskip\bigskip


\begin{raggedright}  

{\it Stefano Perazzini on behalf of the LHCb Collaboration\index{Perazzini, S.}\\
Istituto Nazionale di Fisica Nucleare (INFN)\\
Sezione di Bologna\\
Via Irnerio 46, 40126 Bologna, Italy}
\bigskip\bigskip
\end{raggedright}

\section{Abstract}
In the following we present the measurements of time-dependent $\mathcal{CP}$ violation in charmless $B$ meson decays performed by LHCb analyzing the $p-p$ collision data collected at a center-of-mass energy of 7~TeV during the 2010 and 2011 LHC runs. In particular we will focus on the analysis of charmless two-body $B$ decays where the direct and mixing-induced $CP$ asymmetry terms of the $B^{0}\to\pi^{+}\pi^{-}$ and $B_{s}^{0}\to K^{+}K^{-}$ decays have been measured using 0.69~fb$^{-1}$ of data collected during 2011. The measurement of the branching ratio of the $B_{s}^{0}\to K^{*0}\bar{K}^{*0}$ decay, using 35~pb$^{-1}$ collected during 2010, is also reported. In the end we show the relative branching ratios of all the decay modes of $B_{\left(s\right)}^{0}\to K_S h^{+}h^{\prime -}$ decays (where $h^{\left(\prime\right)}=\pi,K$), measured analyzing 1~fb$^{-1}$ of data collected during 2011.

\section{Introduction}

In this paper we report measurements of time-dependent $\mathcal{CP}$ violation in charmless $B$ meson decays performed  at LHCb~\cite{LHCB}. We will also briefly discuss the LHCb potential with other measurements in this sector. All the results here shown are obtained analyzing the $p-p$ collision data collected at a center-of-mass energy of 7~TeV during the 2010 and 2011 LHC runs. The time-dependent $\mathcal{CP}$ asymmetry of either a $B^0$ or a $B_s^0$ meson decaying into a $\mathcal{CP}$-eigenstate $f$ can be written as:
\begin{equation}
A_{\mathcal{CP}}\left(t\right) = \frac{\Gamma_{\bar{B}\to f}\left(t\right)-\Gamma_{B\to f}\left(t\right)}{\Gamma_{\bar{B}\to f}\left(t\right)+\Gamma_{B\to f}\left(t\right)} = \frac{A_{dir}\cos{\left(\Delta mt\right)}+A_{mix}\sin{\left(\Delta mt\right)}}{\cosh{\left(\frac{\Delta\Gamma}{2}t\right)}+A_{\Delta}\sinh{\left(\frac{\Delta\Gamma}{2}t\right)}},
\end{equation}
where $\Gamma\left(t\right)$ represents the time dependent decay rate of the initial $B$ or $\bar{B}$ meson to the final state $f$, $\Delta m$ and $\Delta\Gamma$ are the $B$ meson oscillation frequency and decay width difference respectively, and where the relation $A_{dir}^{2}+A_{mix}^{2}+A_{\Delta}^{2}=1$ holds. Within this parameterization, $A_{dir}$ and $A_{mix}$ account for $\mathcal{CP}$ violation in the decay and in the interference between mixing and decay, respectively. Since the decay amplitudes of charmless $B$ decays receive important contribution from penguin diagrams, such a class of decays represents a powerful tool in the search for physics beyond the Standard Model. In fact new particles may appear as virtual contributions inside the loop diagrams, altering the Standard Model expectation for $A_{dir}$ and $A_{mix}$.

The LHCb experiment has a great potential in this sector thanks to its capabilities of efficiently triggering and reconstructing charmless $B$ decays, to its excellent decay-time resolution allowing to follow the fast $B_s^0$ oscillations and to the large cross section for $B$ hadron production at LHC.

\section{$B^0\to \pi^+\pi^-$ and $B_s^0\to K^+ K^-$}

The $B^0\to\pi^+\pi^-$ and $B_s^0\to K^+K^-$ decay amplitudes receive contributions from tree, electroweak penguin, strong penguin and annihilation topologies. The presence of penguin diagrams makes these decays sensitive to New Physics, but with the drawback of theoretical uncertainties in the determination of relevant hadronic parameters. In Refs.~\cite{B2HH_THEO} strategies are presented in order to use the $U$-spin symmetry ({\it i.e.} the invariance of strong interaction dynamics under the quark exchange $d\leftrightarrow s$) relating the $B^0\to\pi^+\pi^-$ and $B_s^0\to K^+K^-$ decays, in order to constraint hadronic parameters and determine the CKM phase $\gamma$ and the $B_s^0$ mixing phase $\phi_s$.

Crucial aspects for measuring time dependent $\mathcal{CP}$ asymmetries in $B^0\to\pi^+\pi^-$ and $B_s^0\to K^+K^-$ decays are the event selection, the calibration of the particle identification (PID) and the flavour tagging. These decays are mainly selected by the hadronic trigger of LHCb, that exploits high transverse momentum and large impact parameter with respect to primary vertices, typical of $B$ hadron decay products. Then the sample is further refined applying a set of kinematic and topological criteria. Finally the 8 final states ($\pi^+\pi^-$, $K^+K^-$, $K^+\pi^-$, $\pi^+ K^-$, $p\pi^-$, $\bar{p}\pi^+$, $pK^-$ and $\bar{p}K^+$) are separated into exclusive subsamples by means of the PID information provided by the two ring-imaging Cherenkov detectors of LHCb. The calibration of the PID observables is performed using large samples of pions, kaons and protons selected from $D^{*+}\rightarrow D^{0}(K^{-}\pi^{+})\pi^{+}$ and $\Lambda\rightarrow p\pi^{-}$ decays (and their charge conjugates). As a demonstration of its PID capabilities, using $0.37$~fb$^{-1}$ of integrated luminosity collected during 2011, LHCb measured the relative branching ratios of various two-body charmless $B$ decays~\cite{B2HHBR}. The measurements of the branching ratios of the $B_s^0\to K^+K^-$ and $B_s^0\to \pi^+K^-$ decays are the most precise to date. In addition, the $B_s^0\to\pi^+\pi^-$ decay is observed for the first time ever with a significance of more than 5$\sigma$.

The determination of the initial flavour of the signal $B$ meson (the so-called ''flavour tagging'') is obtained using a multivariate algorithm that analyzes the decay products of the other $B$ hadron in the event~\cite{LHCBTAGGING}. The response of the algorithm is calibrated by measuring the oscillation of the flavour specific decay $B^{0}\rightarrow K^{+}\pi^{-}$, in which the amplitude is related to the effective mistag rate. Using the measured mistag rate as an external input to a two dimensional (invariant mass and decay time) maximum likelihood fit of the $\pi^{+}\pi^{-}$ and $K^{+}K^{-}$ spectra, the direct and mixing-induced $CP$ asymmetry terms $A_{\pi\pi}^{dir}$, $A_{\pi\pi}^{mix}$, $A_{KK}^{dir}$ and $A_{KK}^{mix}$ have been measured. The results, obtained using 0.69~fb$^{-1}$ of data collected during 2011, are~\cite{B2HHTD}:
\begin{eqnarray}
A_{\pi\pi}^{dir} = 0.11 \pm 0.21\mathrm{(stat.)} \pm 0.03\mathrm{(syst.)} \\
A_{\pi\pi}^{mix} = -0.56 \pm 0.17\mathrm{(stat.)} \pm 0.03\mathrm{(syst.)} \\
A_{KK}^{dir} = 0.02 \pm 0.18\mathrm{(stat.)} \pm 0.04\mathrm{(syst.)} \\
A_{KK}^{mix} = 0.17 \pm 0.18\mathrm{(stat.)} \pm 0.05\mathrm{(syst.)},
\end{eqnarray}
with correlations $\rho\left(A_{\pi\pi}^{dir},A_{\pi\pi}^{mix}\right)=-0.34$ and $\rho\left(A_{KK}^{dir},A_{KK}^{mix}\right)=-0.10$. $A_{\pi\pi}^{dir}$ and $A_{\pi\pi}^{mix}$ are measured for the first time at a hadronic machine and are compatible with the previous results from $B$-Factories~\cite{BABARBELLE}. The time-dependent $\mathcal{CP}$ asymmetry terms $A_{KK}^{dir}$ and $A_{KK}^{mix}$ are measured for the first time ever and are compatible with theoretical expectations~\cite{ROADMAP}.

\section{Other charmless decays}

Using other charmless $B$ decays where time-dependent $\mathcal{CP}$ asymmetries can be measured, LHCb already produced results. Using just 35~pb$^{-1}$ collected during 2010, LHCb observed for the first time the $B_{s}^{0}\to K^{*0}\bar{K}^{*0}$ decay~\cite{BSKST}. This decay is governed by pure penguin topologies offering an interesting potential in the quest for New Physics. It has been pointed out~\cite{BSKST_THEO} that the time dependent $\mathcal{CP}$ asymmetries in this decay can be used to measure the CKM phase $\gamma$ and the $B_s^0$ mixing phase $\phi_s$ owing to the information provided by the $U$-spin related decay $B^{0}\to K^{*0}\bar{K}^{*0}$. The signal yield obtained from the fit to the invariant mass distribution is equal to $49.8\pm 7.5$ candidates and the corresponding branching ratio is $\mathcal{BR}\left(B_{s}^{0}\to K^{*0}\bar{K}^{*0}\right) = \left(2.81 \pm 0.46 \mathrm{(stat.)} \pm 0.45 \mathrm{(syst.)} \pm 0.34 \left(f_s/f_d\right)\right)\times10^{-5}$, where the last error comes from the knowledge of the $b$-quark hadronization probabilities. 

\begin{table}[h]
  \begin{center}
    \begin{tabular}{l|c}  
      Decay &  $\mathcal{BR}\times 10^{-6}$ \\  
      \hline
      $\mathcal{BR}\left(B^0\to K_SK^{\pm}\pi^{\mp}\right)$ & $5.8 \pm 0.9 \pm 0.9 \pm 0.2$ \\
      $\mathcal{BR}\left(B^0\to K_SK^{+}K^{-}\right)$ & $26.3 \pm 2.0 \pm 2.0 \pm 1.1$ \\
      $\mathcal{BR}\left(B_s^0\to K_S\pi^{+}\pi^{-}\right)$ & $11.9 \pm 3.0 \pm 2.0 \pm 0.5$ \\
      $\mathcal{BR}\left(B_s^0\to K_SK^{\pm}\pi^{\mp}\right)$ & $97 \pm 7 \pm 10 \pm 4$ \\
      $\mathcal{BR}\left(B_s^0\to K_SK^{+}K^{-}\right)$ & $4.2 \pm 1.5 \pm 0.9 \pm 0.2$ \\
    \end{tabular}
    \caption{Branching ratios of the $B_{\left(s\right)}^{0}\to K_S h^{+}h^{\prime -}$ decays measured by LHCb. The first error is statistical, the second is systematic and the third is due to the uncertainty on $\mathcal{BR}\left(B^0\to K_s\pi^+\pi^-\right)$ that is used as a normalization.}
    \label{tab:B2KSHH}
  \end{center}
\end{table}

Another important result is the measurement of the relative branching ratios of all the decay modes of $B_{\left(s\right)}^{0}\to K_S h^{+}h^{\prime -}$ decays (where $h^{\left(\prime\right)}=\pi,K$)~\cite{B2KSHH}.  The dominant topologies governing these decays are $b\rightarrow q\bar{q}s$ loop transitions (with $q=u,d,s$), where new particles in several extensions of the Standard Model may appear as virtual contributions. Several strategies have been proposed in order to extract information on the CKM phase $\gamma$ and also on the $B^0$ and $B_s^0$ mixing phases $\phi_d$ and $\phi_s$~\cite{B2KSHH_THEO} from the time-dependent analysis of these decays in the Dalitz plane. As a first step, using the full 1~fb$^{-1}$ collected during the 2011, LHCb measured the relative branching ratios of all the $B_{\left(s\right)}^{0}\to K_S h^{+}h^{\prime -}$ modes with respect to the well established $\mathcal{BR}\left(B^0\to K_s\pi^+\pi^-\right)$. The results are reported in Tab.~\ref{tab:B2KSHH} where the first error is statistical, the second is systematic and the third is due to the uncertainty on $\mathcal{BR}\left(B^0\to K_s\pi^+\pi^-\right)$. First evidence of the two decays $B_s^0\to K_S\pi^+\pi^-$ and $B_s^0\to K_S K^+K^-$ is obtained with a significance of 4.3$\sigma$ and 3.3$\sigma$ respectively. Furthermore, LHCb observed for the first time the $B_s^0\to K_S K^{\pm}\pi^{\mp}$ decay.



\end{document}